\documentclass[12pt]{article}
\usepackage{epsfig}
\usepackage{axodraw}

\newcommand\pubnumber{PSI-PR-01-02}
\newcommand\pubdate{\today}
\newcommand\hepnumber{hep-ph/0101213}

\def\Title#1{\begin{center} {\Large\bf #1 } \end{center}}
\def\Author#1{\begin{center}{ \sc #1} \end{center}}
\def\Address#1{\begin{center}{ \it #1} \end{center}}

\newcommand\pubblock{\rightline{\begin{tabular}{l} \pubnumber\\
         \pubdate\\ \hepnumber \end{tabular}}}
\newenvironment{Abstract}{\begin{quotation}  }{\end{quotation}}
\newenvironment{Presented}{\begin{quotation} \begin{center} 
             Presented at the\end{center}
      \begin{center}\begin{large}}{\end{large}\end{center} \end{quotation}}
\def\Acknowledgments{\bigskip  \bigskip \begin{center}
          \large\bf Acknowledgments\end{center}}

\makeatletter
\def\section{\@startsection{section}{0}{\z@}{5.5ex plus .5ex minus
 1.5ex}{2.3ex plus .2ex}{\large\bf}}
\def\subsection{\@startsection{subsection}{1}{\z@}{3.5ex plus .5ex minus
 1.5ex}{1.3ex plus .2ex}{\normalsize\bf}}
\def\subsubsection{\@startsection{subsubsection}{2}{\z@}{-3.5ex plus
-1ex minus  -.2ex}{2.3ex plus .2ex}{\normalsize\sl}}

\renewcommand{\@makecaption}[2]{%
   \vskip 10pt
   \setbox\@tempboxa\hbox{\small #1: #2}
   \ifdim \wd\@tempboxa >\hsize     % IF longer than one line:
       \small #1: #2\par          %   THEN set as ordinary paragraph.
     \else                        %   ELSE  center.
       \hbox to\hsize{\hfil\box\@tempboxa\hfil}
   \fi}

 \def\citenum#1{{\def\@cite##1##2{##1}\cite{#1}}}
 
\newcount\@tempcntc
\def\@citex[#1]#2{\if@filesw\immediate\write\@auxout{\string\citation{#2}}\fi
  \@tempcnta\z@\@tempcntb\m@ne\def\@citea{}\@cite{\@for\@citeb:=#2\do
    {\@ifundefined
       {b@\@citeb}{\@citeo\@tempcntb\m@ne\@citea\def\@citea{,}{\bf ?}\@warning
       {Citation `\@citeb' on page \thepage \space undefined}}%
    {\setbox\z@\hbox{\global\@tempcntc0\csname b@\@citeb\endcsname\relax}%
     \ifnum\@tempcntc=\z@ \@citeo\@tempcntb\m@ne
       \@citea\def\@citea{,}\hbox{\csname b@\@citeb\endcsname}%
     \else
      \advance\@tempcntb\@ne
      \ifnum\@tempcntb=\@tempcntc
      \else\advance\@tempcntb\m@ne\@citeo
      \@tempcnta\@tempcntc\@tempcntb\@tempcntc\fi\fi}}\@citeo}{#1}}
\def\@citeo{\ifnum\@tempcnta>\@tempcntb\else\@citea\def\@citea{,}%
  \ifnum\@tempcnta=\@tempcntb\the\@tempcnta\else
  {\advance\@tempcnta\@ne\ifnum\@tempcnta=\@tempcntb \else\def\@citea{--}\fi
    \advance\@tempcnta\m@ne\the\@tempcnta\@citea\the\@tempcntb}\fi\fi}
\makeatother

\def\beq{\begin{equation}}
\def\eeq#1{\label{#1}\end{equation}}
\def\eeqn{\end{equation}}

\newenvironment{Eqnarray}%
   {\arraycolsep 0.14em\begin{eqnarray}}{\end{eqnarray}}
\def\beqa{\begin{Eqnarray}}
\def\eeqa#1{\label{#1}\end{Eqnarray}}
\def\eeqan{\end{Eqnarray}}

\let\littlebar=\bar
\let\bar=\overbar

\def\ie{{\it i.e.}}

\def\M{{\cal M}}
\def\O{{\cal O}}

\def\Dslash{\not{\hbox{\kern-4pt $D$}}}
\def\dslash{\not{\hbox{\kern-2pt $\del$}}}

\def\eff{{\mbox{\scriptsize eff}}}

\def\gt{\Gamma_t}

\def\msb{{\bar{\ssstyle M \kern -1pt S}}}

\def\lsim{\mathrel{\raise.3ex\hbox{$<$\kern-.75em\lower1ex\hbox{$\sim$}}}}
\def\gsim{\mathrel{\raise.3ex\hbox{$>$\kern-.75em\lower1ex\hbox{$\sim$}}}}

\makeatletter

\expandafter\ifx\csname mathrm\endcsname\relax\def\mathrm#1{{\rm #1}}\fi

\newcounter{saveeqn}

\@addtoreset{equation}{section}

\def\beq{\begin{equation}}
\def\eeq{\end{equation}}
\def\beqar{\begin{eqnarray}}
\def\eeqar{\end{eqnarray}}
\def\barr#1{\begin{array}{#1}}
\def\earr{\end{array}}
\def\bfi{\begin{figure}}
\def\efi{\end{figure}}
\def\btab{\begin{table}}
\def\etab{\end{table}}
\def\bce{\begin{center}}
\def\ece{\end{center}}
\def\nn{\nonumber}
\def\nl{\nonumber\\}

\def\al{\alpha}

\def\de{\delta}
\def\la{\lambda}
\def\si{\sigma}

\def\refeq#1{\mbox{(\ref{#1})}}
\def\reffi#1{\mbox{Figure~\ref{#1}}}
\def\reffis#1{\mbox{Figures~\ref{#1}}}

\def\refse#1{\mbox{Section~\ref{#1}}}
\def\refses#1{\mbox{Sections~\ref{#1}}}

\def\citere#1{\mbox{Ref.~\cite{#1}}}
\def\citeres#1{\mbox{Refs.~\cite{#1}}}

\newcommand{\TeV}{\unskip\,\mathrm{TeV}}
\newcommand{\GeV}{\unskip\,\mathrm{GeV}}

\newcommand{\ri}{{\mathrm{i}}}
\newcommand{\rd}{{\mathrm{d}}}

\newcommand{\rR}{{\mathrm{R}}}
\newcommand{\rT}{{\mathrm{T}}}
\newcommand{\rL}{{\mathrm{L}}}

\renewcommand{\O}{{\cal O}}

\def\mathswitchr#1{\relax\ifmmode{\mathrm{#1}}\else$\mathrm{#1}$\fi}
\newcommand{\PW}{\mathswitchr W}

\newcommand{\PZ}{\mathswitchr Z}
\newcommand{\PA}{\mathswitchr A}
\newcommand{\PH}{\mathswitchr H}
\newcommand{\Pf}{\mathswitch f}
\newcommand{\Pfbar}{\mathswitch \littlebar{f}}
\newcommand{\Pe}{\mathswitchr e}

\newcommand{\Pb}{\mathswitchr b}
\newcommand{\Pt}{\mathswitchr t}
\newcommand{\Pep}{\mathswitchr {e^+}}
\newcommand{\Pem}{\mathswitchr {e^-}}
\newcommand{\PWp}{\mathswitchr {W^+}}
\newcommand{\PWm}{\mathswitchr {W^-}}

\def\mathswitch#1{\relax\ifmmode#1\else$#1$\fi}
\newcommand{\MW}{\mathswitch {M_\PW}}
\newcommand{\MZ}{\mathswitch {M_\PZ}}
\newcommand{\MH}{\mathswitch {M_\PH}}

\newcommand{\Mt}{\mathswitch {m_\Pt}}

\newcommand{\cw}{\mathswitch {c_\mathrm{w}}}
\newcommand{\sw}{\mathswitch {s_\mathrm{w}}}
\newcommand{\NCf}{\mathswitch {N_{\mathrm{C}}^f}}

\newcommand{\SU}{\mathrm{SU}}

\newcommand{\SUtwo}{\mathrm{SU(2)}}
\newcommand{\Uone}{\mathrm{U}(1)}

\def\ie{i.e.\ }

\def\cf{cf.\ }

\newcommand{\elm}{{\mathrm{em}}}
\newcommand{\ew}{{\mathrm{ew}}}
\newcommand{\sew}{{\mathrm{ew}}}

\newcommand{\coll}{{\mathrm{coll}}}

\newcommand{\SC}{{\mathrm{LSC}}}
\renewcommand{\SS}{{\mathrm{SSC}}}
\newcommand{\cc}{{\mathrm{C}}}

\newcommand{\pre}{{\mathrm{PR}}}
\newcommand{\Yuk}{{\mathrm{Yuk}}}

\newcommand{\bew}{b^{\ew}}

\newcommand{\cew}{C^{\ew}}

\newcommand{\NB}{N}
\newcommand{\GB}{V}

\newcommand{\ls}{l(s)}
\newcommand{\lu}{l(\mu^2)}

\newcommand{\lsl}{l_{\cc}}
\newcommand{\lpr}{l_{\pre}}
\newcommand{\lYuk}{l_{\Yuk}}

\newcommand{\lemf}{l^\elm(m_f^2)}

\newcommand{\lemfsi}{l^\elm(m_{f_\si}^2)}

\newcommand{\lemphi}{l^\elm(m_{\varphi}^2)}

\newcommand{\lemW}{l^\elm(\MW^2)}
\newcommand{\Ls}{L(s)}

\newcommand{\Lemk}{L^\elm(s,\lambda^2,m_k^2)}
\newcommand{\Lemphi}{L^\elm(s,\lambda^2,m_\varphi^2)}

\newcommand{\lrs}{\log{\frac{|r_{kl}|}{s}}}

\newcommand{\ltu}{\log{\frac{t}{u}}}
\newcommand{\lts}{\log{\frac{|t|}{s}}}

\def\draftdate{\relax}
\def\mda{\relax}
\def\mua{\relax}
\def\mla{\relax}
\def\draft{
\def\thtystars{******************************}
\def\sixtystars{\thtystars\thtystars}
\typeout{}
\typeout{\sixtystars**}
\typeout{* Draft mode!
         For final version remove \protect\draft\space in source file *}
\typeout{\sixtystars**}
\typeout{}
\def\draftdate{\today}
\def\mua{\marginpar[\boldmath\hfil$\uparrow$]%
                   {\boldmath$\uparrow$\hfil}%
                    \typeout{marginpar: $\uparrow$}\ignorespaces}
\def\mda{\marginpar[\boldmath\hfil$\downarrow$]%
                   {\boldmath$\downarrow$\hfil}%
                    \typeout{marginpar: $\downarrow$}\ignorespaces}
\def\mla{\marginpar[\boldmath\hfil$\rightarrow$]%
                   {\boldmath$\leftarrow $\hfil}%
                    \typeout{marginpar: $\leftrightarrow$}\ignorespaces}
\def\Mua{\marginpar[\boldmath\hfil$\Uparrow$]%
                   {\boldmath$\Uparrow$\hfil}%
                    \typeout{marginpar: $\uparrow$}\ignorespaces}
\def\Mda{\marginpar[\boldmath\hfil$\Downarrow$]%
                   {\boldmath$\Downarrow$\hfil}%
                    \typeout{marginpar: $\downarrow$}\ignorespaces}
\def\Mla{\marginpar[\boldmath\hfil$\Rightarrow$]%
                   {\boldmath$\Leftarrow $\hfil}%
                    \typeout{marginpar: $\leftrightarrow$}\ignorespaces}
\def\sua{\Mua}
\def\sda{\Mda}
\def\sla{\Mla}
\overfullrule 5pt
\oddsidemargin -15mm
\marginparwidth 29mm
}

\begin{document}
\begin{titlepage}
\pubblock

\vfill
\def\thefootnote{\fnsymbol{footnote}}
\Title{Leading electroweak logarithms\\[5pt] at one loop}
\vfill
\Author{Ansgar Denner}
\Address{Paul Scherrer Institut,
CH-5232 Villigen PSI, Switzerland}
\Author{Stefano Pozzorini}
\Address{Institute of Theoretical Physics, University of Z\"urich, Switzerland \\[1ex]and \\[1ex]
Paul Scherrer Institut,
CH-5232 Villigen PSI, Switzerland}
\vfill
\begin{Abstract}
We summarize results for the complete one-loop electroweak logarithmic
corrections for general processes at high energies and fixed angles.
Our results are applicable to arbitrary matrix elements that are not
mass-suppressed.  We give explicit results for $\PW$-boson-pair production in $\Pep\Pem$ annihilation.
\end{Abstract}
\vfill
\begin{Presented}
5th International Symposium on Radiative Corrections \\ 
(RADCOR--2000) \\[4pt]
Carmel CA, USA, 11--15 September, 2000
\end{Presented}
\vfill
\end{titlepage}
\def\thefootnote{\arabic{footnote}}
\setcounter{footnote}{0}

\section{Introduction}
Future colliders, such as the LHC \cite{cernreport} 
or an $\Pep\Pem$ linear collider (LC) \cite{LC1},
will explore the energy range %above the electrowak scale, 
$\sqrt{s}\gg \MZ$.  It is known since
many years (see, for instance, \citeres{Kuroda,eeWWhe}) that above the
electroweak scale the structure of the leading electroweak corrections
changes and double logarithms of Sudakov type \cite{SUD}  as well
as single logarithms involving the ratio of the energy to the
electroweak scale become dominating.
These logarithms arise from virtual (or real) gauge bosons emitted by
the initial and final-state particles. They correspond to the
well-known soft and collinear singularities observed in 
QCD. 

In the electroweak theory, unlike in massless gauge theories,
the large logarithms originating from virtual corrections are of physical significance. In fact%, unlike for the photon
, real $\PZ$-boson and $\PW$-boson bremsstrahlung need not be included,
since the masses of the weak gauge bosons, $\PZ$ and $\PW$, provide
a physical cutoff, and the massive gauge bosons can be detected
as distinguished particles.

The typical size of double-logarithmic (DL) and single-logarithmic
(SL) corrections is given by
\beq\label{largelogs}%\refeq{largelogs}
\frac{\alpha}{4\pi \sw^2}\log^2{\frac{s}{\MW^2}}= 6.6 \%,\qquad \frac{\alpha}{4\pi \sw^2}\log{\frac{s}{\MW^2}}=1.3\%
\eeq
at $\sqrt{s}=1\TeV$ and increases with energy. If the experimental
precision is at the few-percent level like at the LHC, both DL and SL
electroweak contributions have to be included at the one-loop level. In view of
the precision objectives of a LC, between the percent and
the permil level, besides the complete one-loop
corrections also two-loop DL effects have to be taken into account.

Owing to this phenomenological relevance, the infrared (IR) 
structure of the
electroweak theory is receiving increasing interest recently.  The
one-loop structure and the origin of the DL corrections have been
discussed for $\Pep\Pem\to\Pf\Pfbar$ \cite{CC0,Ku1} and are by
now well established. Recipes for the resummation of the DL
corrections have been developed
\cite{CC1,Ku1,Ku2,Fa00} and explicit calculations of the leading DL
corrections for the processes $g\to\Pf\Pfbar$ and
$\Pep\Pem\to\Pf\Pfbar$ have been performed \cite{Me2,Be00,Ho00}.
On the other hand, for the SL corrections complete one-loop
calculations are only available for 4-fermion
neutral-current processes \cite{be1,Ku2} and \PW-pair production 
\cite{eeWWhe}. The subleading two-loop logarithmic corrections have
been evaluated for $\Pep\Pem\to\Pf\Pfbar$ in \citere{Ku2}. A general
recipe for a subclass of SL corrections to all orders has been proposed in
\citere{Me1}, based on the infrared-evolution-equation method.
 
Here, we summarize the results for all DL and SL contributions to
the electroweak one-loop virtual corrections published in \citere{DennPozz1}.
The results apply to
exclusive processes with arbitrary external states, including
transverse and longitudinal gauge bosons as well as Higgs fields.

The paper is organized as follows: in \refse{se:defcon} we introduce
our notations and discuss the origin of the leading electroweak
logarithms. The leading logarithms originating from the
soft--collinear region, from the soft or collinear regions, and from
parameter renormalization are considered in \refses{se:soft-coll},
\ref{se:soft-or-coll}, and \ref{se:ren}, respectively. In
\refse{se:applicat} we apply our general results to W-boson-pair
production in $\Pep\Pem$ annihilation.

\section{Form and origin of enhanced logarithmic corrections}
\label{se:defcon} 
We consider electroweak processes involving $n$ arbitrary incoming%
\footnote{Usual scattering processes are obtained by crossing symmetry}
particles (or antiparticles) associated to the fields $\varphi_{i_k}$, 
\beq \label{process}%\refeq{process}
\varphi_{i_1}(p_1)\dots \varphi_{i_n}(p_n)\rightarrow 0.
\eeq
The  indices  $i_k$ correspond to the reducible representation of $\SUtwo\times\Uone$ including all fields in the standard model, and we restrict ourselves to
Born matrix elements
$
\M_0^{i_1 \ldots i_n}(p_1,\ldots, p_n)
$
that are not suppressed in the limit where
all invariants are much larger than the gauge-boson masses,
\beq \label{Sudaklim} % \refeq{Sudaklim}
r_{kl}:=(p_k+p_l)^2\sim 2p_kp_l \gg \MW^2.
\eeq
In the high-energy limit \refeq{Sudaklim}, we split all  enhanced DL and SL corrections into a ``symmetric electroweak'' (ew) 
part given by logaritms of the ratio between the energy and the electroweak scale \refeq{largelogs}
and a remaining part that we denote as ``pure electromagnetic contribution'' (em), which involves logarithms of the  light-fermion masses and the infinitesimal photon mass $\la$ used to regularize IR singularities.
For the symmetric electroweak logarithms we introduce the shorthands
\beq \label{dslogs}%\refeq{dslogs}
\Ls:=\frac{\alpha}{4\pi}\log^2{\frac{s}{\MW^2}},\qquad
\ls:=\frac{\alpha}{4\pi}\log{\frac{s}{\MW^2}}.
\eeq
We assume that  the masses $\MH$, $\Mt$, $\MZ$, and $\MW$ have the
same order of magnitude and neglect all logarithms of ratios of these
masses.

In logarithmic approximation (LA)
the one-loop corrections  to \refeq{process} assume the form  
\beq \label{LAfactorization} %\refeq{LAfactorization}
\delta \M^{i_1 \ldots i_n}(p_1,\ldots, p_n)= 
\M_0^{i'_1 \ldots i'_n}(p_1, \ldots, p_n)\delta_{i'_1i_1 \ldots i'_ni_n},
\end{equation}
\ie they factorize into the lowest-order matrix element 
times an $\SUtwo\times\Uone$ matrix.
For matrix elements that are not mass-suppressed the factorization
formula is universal. The matrix $\delta_{i'_1i_1 \ldots i'_ni_n}$ can
be expressed using the couplings $\ri eI^{\GB_a}(\varphi)$ of the
external fields $\varphi_{i_k}$ to the gauge bosons $\GB_a$. These
correspond to the generators of infinitesimal global
$\SUtwo\times\Uone$ transformations of these fields,% 
\footnote{Details about the explicit form of the generators and other
  group theoretical quantities can be found in the appendix of
  \citere{DennPozz1}.}
\beq \label{generators}%\refeq{generators}
\delta_{\GB_a} \varphi_{i}=
\ri e I^{\GB_a}_{\varphi_i\varphi_{i'}}(\varphi)\,\varphi_{i'}.
\eeq
In terms of the electric charge $Q$ and weak isospin $T^a$ they are given by
\beq
I^A=-Q,\qquad I^Z=\frac{T^3-\sw^2 Q}{\sw\cw},\qquad
I^\pm=%\frac{1}{\sw}T^\pm=
\frac{T^1\pm\ri T^2}{\sqrt{2}\sw}
\eeq
and depend on the weak mixing angle, which is fixed 
by $\cw^2=1-\sw^2=\MW^2/\MZ^2$.  

In general, large logarithms contributing to \refeq{LAfactorization}
are shared between %the mass-singular 
the loop diagrams and the coupling- and field-renormalization
constants, depending on the gauge-fixing and the renormalization
scheme.  We work within the 't~Hooft--Feynman gauge and adopt the
on-shell scheme \cite{FortPhys} for field and parameter
renormalization.  
We use dimensional regularization
and  choose the regularization scale $\mu^2=s$ so that the logarithms
$\log{(\mu^2/s)}$ 
related to the UV singularities are not enhanced, and only the
mass-singular logarithms $\log{(\mu^2/M^2)}$ or $\log{(s/M^2)}$ are
large. 
In this setup large logarithms are distributed as follows:
\begin{itemize}
\item The DL contributions originate from those one-loop diagrams
  where soft--colline\-ar gauge bosons are exchanged between pairs of
  external legs. These double logarithms are obtained with the eikonal
  approximation.
\item The SL mass-singular contributions from loop diagrams originate
  from the emission of virtual collinear gauge bosons from external
  lines \cite{KLN}. These SL contributions are extracted from the loop
  diagrams in the collinear limit by means of Ward identities, and are
  found to factorize into the Born amplitude times ``collinear
  factors''  \cite{DennPozz2} . %These are the main result of this article,
  
\item The remaining SL contributions originating from soft and
  collinear regions are contained in the field renormalization
  constants (FRCs).
\item The parameter renormalization (PR) constants, \ie the charge- and
  weak-mixing-angle renormalization constants, as well as the
  renormalization of dimensionless mass ratios associated with the
  Yukawa and the scalar self-couplings, involve the SL contributions
  of UV origin. These are the logarithms that are controlled by 
  the renormalization group.
\end{itemize}
The DL and SL mass-singular terms are extracted from loop diagrams by
setting all masses to zero in the numerators of the loop-integrals.
For processes  
involving external longitudinal gauge bosons, this approach is not directly applicable, owing to the longitudinal polarization vectors 
\beq \label{longplovec}%\refeq{longplovec}
\epsilon_\rL^\mu(p)=\frac{p^\mu}{M}+\O\left(\frac{M}{p^0}\right),
\eeq 
which are inversely proportional to the gauge boson mass. However,
since we are only interested in the high-energy limit, we can use the
Goldstone-boson equivalence theorem \cite{et} taking into account the
correction factors from higher-order contributions \cite{etcorr}.

\section{Soft--collinear contributions}
\label{se:soft-coll}

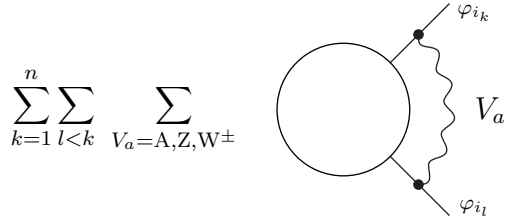
\begin{figure}[b!] 
\begin{center}
\begin{picture}(140,80)(0,10)
\Line(70,50)(110,90)
\Line(70,50)(110,10)
\PhotonArc(70,50)(40,-45,45){2}{5}
\GCirc(70,50){25}{1}
\Text(30,50)[r]{${\displaystyle \sum\limits_{k=1}^n\sum\limits_{l<k}\,\,\sum\limits_{\GB_a=\PA,\PZ,\PW^\pm}}$}
\Text(120,50)[l]{$\GB_a$}
\Text(120,90)[t]{\scriptsize$\varphi_{i_k}$}
\Text(120,10)[b]{\scriptsize$\varphi_{i_l}$}
\Vertex(98.3,78.3){2}
\Vertex(98.3,21.7){2}
\end{picture}
\end{center}
\caption{Feynman diagrams leading to DL corrections}
\label{SCdiag}
\end{figure}

The DL corrections originate from loop diagrams where virtual gauge
bosons $\GB_a=\PA,\PZ,\PW^\pm$ 
are exchanged between pairs of external
legs (\reffi{SCdiag}), 
and arise from the integration region where the
gauge-boson momenta are soft and collinear to one of the external
legs.
They are obtained using the eikonal
approximation, and result in a double sum over pairs of external legs
\beqar \label{eikonalapp}%\refeq{eikonalapp}
\delta^{\mathrm{DL}} \M^{i_1 \ldots i_n}
&=&\frac{\alpha}{8\pi}\sum_{k=1}^n\sum_{l\neq k}\sum_{\GB_a=A,Z,W^\pm} I^{\GB_a}_{i'_k i_k}(k) I^{\bar{\GB}_a}_{i'_l i_l}(l) \M_0^{i_1 \ldots i'_k \ldots i'_l \ldots i_n}
\nl&&\times
\left[\log^2\left(\frac{|r_{kl}|}{M^2_{V_a}}\right)
-\de_{\GB_aA}\log^2\left(\frac{m_k^2}{\la^2}\right) \right], 
\eeqar
where $\bar{V}_a$ represents the charge conjugated of $V^a$.  Formula
\refeq{eikonalapp} applies to chiral fermions, Higgs bosons, and
transverse gauge bosons and depends on their gauge couplings
$I^{\GB_a}(k)$.  The DL corrections for external longitudinal gauge
bosons $\PZ_\rL$ and $\PW_\rL^\pm$ are obtained from the corrections
\refeq{eikonalapp} for the corresponding external would-be Goldstone bosons
$\chi$ and $\phi^\pm$, respectively, using the equivalence theorem
\beqar \label{DLeqtheor}%\refeq{DLeqtheor}
\de^{\mathrm{DL}}\M^{\ldots \PW^\pm_\rL \ldots} &=&\de^{\mathrm{DL}}\M^{\ldots \phi^\pm \ldots},\nl
\de^{\mathrm{DL}}\M^{\ldots \PZ_\rL \ldots} &=&\ri\de^{\mathrm{DL}}\M^{\ldots \chi \ldots}.
\eeqar

\subsection*{Leading soft--collinear contributions}
The DL term in \refeq{eikonalapp} containing the invariant $r_{kl}$ depends on the angle between the momenta $p_k$ and $p_l$. Writing
\beq \label{angsplit}%\refeq{angsplit}
\log^2\left(\frac{|r_{kl}|}{M^2}\right)=
\log^2\left(\frac{s}{M^2}\right)+
2\log\left(\frac{s}{M^2}\right)
\log\left(\frac{|r_{kl}|}{s}\right)+
\log^2\left(\frac{|r_{kl}|}{s}\right),
\eeq
one can isolate an angular-independent part proportional to $\Ls$, and
this part, together with the additional contributions from photon
loops in \refeq{eikonalapp}, gives the leading soft--collinear ($\SC$)
contribution.  Using the invariance of the $S$ matrix with respect to
global $\SUtwo\times\Uone$ transformations,
the $\SC$ contribution in \refeq{eikonalapp} can be written as a
single sum over external legs,
\beq \label{SCsum}%\refeq{SCsum}
\de^{\SC} \M^{i_1 \ldots i_n} =\sum_{k=1}^n \delta^\SC_{i'_ki_k}(k)
\M_0^{i_1 \ldots i'_k\ldots i_n},
\end{equation} 
where the
correction factors reads
\beq \label{deSC} %\refeq{deSC}
\de^\SC_{i'_ki_k}(k)=- \frac{1}{2}\left[ C^{\ew}_{i'_ki_k}(k)\Ls 
+\de_{i'_ki_k} Q_k^2\Lemk \right].
\eeq
The first term represents the DL symmetric-electroweak part and is
proportional to the effective electroweak Casimir operator%
\footnote{ As explained in \citere{DennPozz1} care must be taken for reducible representations, where owing to mixing, \refeq{CasimirEW} can be non-diagonal.}
\begin{equation}\label{CasimirEW} %\refeq{CasimirEW} 
\cew:=\sum_{\GB_a=A,Z,W^\pm} I^{\GB_a}I^{\bar{\GB}_a}=\frac{1}{\cw^2}\left(\frac{Y}{2}\right)^2+\frac{1}{\sw^2}C^{\SUtwo},
\end{equation}
which depends on the weak hypercharge $Y=2(Q-T^3)$ and the $\SUtwo$ Casimir operator $C^{\SUtwo}$.
The second term in \refeq{deSC}
originates from photon loops and reads
\beq
\Lemk:= 2\ls\log{\left(\frac{\MW^2}{\la^2} \right)}+
\frac{\alpha}{4\pi}\left[\log^2{\frac{\MW^2}{\la^2}}
-\log^2{\frac{m_k^2}{\la^2}}\right].
\eeq

\subsection*{Subleading soft--collinear contributions}
The remaining part of \refeq{eikonalapp} is a subleading
soft--collinear ($\SS$) contribution,
\beq \label{SScorr}%\refeq{SScorr}
\de^\SS \M^{i_1 \ldots i_n} =\sum_{k=1}^n
\sum_{l<k}\sum_{\GB_a=A,Z,W^\pm}\delta^{\GB_a,\SS}_{i'_ki_k i'_li_l}(k,l)
\M_0^{i_1\ldots i'_k\ldots i'_l\ldots i_n}.
\eeq
This remains a  sum over pairs of external legs with angular-dependent
factors% 
\footnote{Double logarithms of $r_{kl}/s$ are neglected in the limit
  \refeq{Sudaklim}.}
\beq \label{subdl1} %\refeq{subdl1}
\de^{V_a,\SS}_{i'_ki_k i'_li_l}(k,l)=
\left[2\ls+\de_{V_aA}\frac{\alpha}{2\pi}\log{\frac{\MW^2}{\la^2}}\right]\lrs I_{i'_ki_k}^{V_a}(k)I_{i'_li_l}^{\bar{V}_a}(l).
\eeq
Owing to the non-diagonal matrices $I^\pm(k)$ (\cf appendix of
\citere{DennPozz1}), the exchange of soft charged gauge bosons
involves $\SUtwo$-transformed Born matrix elements on the 
right-hand side of \refeq{SScorr}.

\section{Collinear and soft single logarithms}
\label{se:soft-or-coll}
The collinear and soft SL corrections originate from field
renormalization and from mass-singular loop diagrams.
On one hand the FRCs give the well-known factors $\de Z_{\varphi}/2$
for each external leg, containing collinear as well as soft SL
contributions.  On the other hand, mass-singular logarithms arise from
the collinear limit of loop diagrams where an external line splits
into two internal lines \cite{KLN}, one of these internal lines being
a virtual gauge boson $\PA,\PZ,$ or $\PW$.
Both contributions factorize as a sum over the external legs,
\beq\label{subllogfact}%\refeq{subllogfact}
\de^{\cc} \M^{i_1 \ldots i_n} =\sum_{k=1}^n \delta^\cc_{i'_ki_k}(k)
\M_0^{i_1 \ldots i'_k \ldots i_n}
\eeq
with 
\beq\label{subllogfact2}%\refeq{subllogfact2}
 \delta^\cc_{i'_ki_k}(k)=\left.\frac{1}{2}\delta Z^\varphi_{i'_k i_k}+\delta^\coll_{i'_ki_k}(k)\right|_{\mu^2=s}.
\eeq
The factorization of the mass-singular loop diagrams will be presented
in a forthcoming publication \cite{DennPozz2}. Therein, we derive the
factorization identities
 %(\reffi{Colldiag}), 
\beqar\label{collfactorization}%\refeq{collfactorization}
\lefteqn{
\sum_{\GB_a=A,Z,W^\pm} \left\{
\vcenter{\hbox{\begin{picture}(100,60)(0,-30)
\Line(5,0)(25,0)
\Line(25,0)(60.9,14.1)
\Photon(25,0)(55.9,-13.4){-2}{3}
\Vertex(25,0){2}
\GCirc(65,0){15}{1}
\Text(40,-18)[r]{$\GB_a$}
\Text(5,5)[lb]{\scriptsize $\varphi_{i_k}$}
\end{picture}}}
-
\vcenter{\hbox{\begin{picture}(100,80)(-10,-40)
\Line(0,0)(65,0)
\PhotonArc(23,0)(12,180,0){1.5}{3.5}
\Vertex(35,0){2}
\Vertex(11,0){2}
\GCirc(65,0){15}{1}
\Text(23,-18)[t]{$\GB_a$}
\Text(6,5)[cb]{\scriptsize$\varphi_{i_k}$}
\end{picture}}}\right.
}\qquad\\
&& \left.\left. - \sum_{l\neq k}\left[
\vcenter{\hbox{\begin{picture}(100,80)(0,-40)
\Line(65,0)(25,30)
\Line(65,0)(25,-30)
\PhotonArc(65,0)(32.5,143.13,216.87){2}{4.5}
\Vertex(39,19.5){2}
\Vertex(39,-19.5){2}
\GCirc(65,0){15}{1}
\Text(30,0)[r]{$\GB_a$}
\Text(20,27)[rb]{\scriptsize$\varphi_{i_k}$}
\Text(20,-27)[rt]{\scriptsize$\varphi_{i_l}$}
\end{picture}}}
\right]_{\mathrm{eik.~appr.}}\right\}\right|_{\mathrm{coll.}}
=
\sum_{\varphi_{i'_k}}
\vcenter{\hbox{\begin{picture}(80,80)(0,-40)
\Line(55,0)(10,0)
\GCirc(55,0){15}{1}
\Text(10,5)[lb]{\scriptsize$\varphi_{i'_k}$}
\end{picture}}}
\de^\coll_{\varphi_{i'_k}\varphi_{i_k}}.\nn
\eeqar
for fermions, gauge bosons and scalar bosons.  These identities are
obtained by evaluation of the loop diagrams involving the
collinear splitting processes $\varphi_{i_k}(p)\rightarrow
V^a(q)\varphi_{i'_k}(p-q)$, after subtraction of the contributions
already contained in the FRCs and the soft collinear corrections.  In
the limit of collinear gauge-boson emission, the left-hand side of
\refeq{collfactorization} is proportional to
\beq\label{collwiA}%\refeq{collwiA}
\sum_{\GB_a,\varphi_{i'_k}}
\int\!\frac{\rd q^D}{(2\pi)^D}
\frac{-\ri e I^{\bar{V}^a  }_{\varphi_{i'_k}\varphi_{i_k}} q^\mu}{(q^2-M_{V_a}^2)[(p-q)^2-M_{\varphi_{i'_k}}^2]}
\times
\left\{
\vcenter{\hbox{\begin{picture}(85,80)(12,20)
\Line(55,60)(15,60)
\Text(35,65)[b]{\scriptsize $\varphi_{i'_k}(p-q)$}
\Photon(30,30)(70,60){1}{6}
\Text(55,35)[t]{\scriptsize $V^a_\mu(q)$}
\GCirc(70,60){15}{1}
\end{picture}}}
-
\vcenter{\hbox{\begin{picture}(85,80)(2,20)
\Line(55,60)(35,60)%\bar{u}^a(q)
\Line(35,60)(10,60)
\Text(30,65)[b]{\scriptsize $ \varphi_{i'_k}(p-q)$}
\Photon(10,30)(35,60){1}{5}
\GCirc(35,60){1}{0}
\Text(35,40)[t]{\scriptsize $V^a_\mu(q)$}
\GCirc(70,60){15}{1}
\end{picture}}}
\right\},
\eeq
where the diagrams between the curly brackets are contracted with the
gauge-boson momentum $q^\mu$.  These contractions can be simplified
using Ward identities resulting from the BRS symmetry of the
spontaneously broken $\SUtwo\times\Uone$ gauge theory
(\cf\citere{DennPozz2}), and in the limit where $q^\mu$ becomes
collinear to the external momentum $p^\mu$ we obtain
\beq\label{collwi}%\refeq{collwi}
\lim_{q^\mu\rightarrow xp^\mu} q^\mu\times
\left\{
\vcenter{\hbox{\begin{picture}(85,80)(10,20)
\Line(55,60)(15,60)
\Text(35,65)[b]{\scriptsize $\varphi_{i'}(p-q)$}
\Photon(30,30)(70,60){1}{6}
\Text(55,35)[t]{\scriptsize $V^a_\mu(q)$}
\GCirc(70,60){15}{1}
\end{picture}}}
-
\vcenter{\hbox{\begin{picture}(90,80)(0,20)
\Line(55,60)(35,60)%\bar{u}^a(q)
\Line(35,60)(10,60)
\Text(30,65)[b]{\scriptsize $ \varphi_{i'}(p-q)$}
\Photon(10,30)(35,60){1}{5}
\GCirc(35,60){1}{0}
\Text(35,40)[t]{\scriptsize $V^a_\mu(q)$}
\GCirc(70,60){15}{1}
\end{picture}}}
\right\}
=\sum_{\varphi_{i''}}
\vcenter{\hbox{\begin{picture}(80,80)(0,-40)
\Line(55,0)(10,0)
\GCirc(55,0){15}{1}
\Text(10,5)[lb]{\scriptsize $\varphi_{i''}(p)$}
\end{picture}}}
 e I^{V^a}_{\varphi_{i''}\varphi_{i'}},
\eeq
up to mass-suppressed terms. Combining \refeq{collwi} with
\refeq{collwiA} we obtain the factorization identity \refeq{collfactorization} 
with the collinear factors
\beq
\de^\coll_{\varphi_{i'}\varphi_{i}}=
\frac{\alpha}{4\pi}K\left[\cew_{\varphi_{i'}\varphi_{i}}\log{\frac{\mu^2}{\MW^2}}+ \de_{\varphi_{i'}\varphi_{i}}Q^2_{\varphi_{i}}\log{\frac{\MW^2}{M_{\varphi_i}^2}} \right],
\eeq
where $K=2$ for fermions and $K=1$ for Higgs bosons, would-be Goldstone bosons,
and gauge bosons.

In the following, we present the complete SL corrections
\refeq{subllogfact2} for the cases of external fermions, transverse
and longitudinal gauge bosons, and Higgs bosons.

\subsection*{Chiral fermions}
For fermions $f^\kappa_\si$ with chirality $\kappa=\rR,\rL$ and
isospin indices $\si=\pm$
\beq \label{deccfer}%\refeq{deccfer}
\de^{\cc}_{f_\si f_{\si'}}(f^\kappa)=\de_{\si\si'}\left\{\left[\frac{3}{2} \cew_{f^\kappa} -\frac{1}{8\sw^2}\left((1+\delta_{\kappa \rR})\frac{m_{f_\si}^2}{\MW^2}+\delta_{\kappa \rL}\frac{m_{f_{-\si}}^2}{\MW^2}\right)\right]\ls+Q_{f_\si}^2\lemfsi\right\}.
\eeq
Besides the contribution of the Casimir operator \refeq{CasimirEW}, we
have Yukawa terms proportional to the masses of the fermion $f_\si$
and of its isospin partner $f_{-\si}$. These are large for
$f^\kappa_\si=\Pt^\rR$, $\Pt^\rL$, and $\Pb^\rL$, where one of the
masses is $\Mt$.  The pure electromagnetic logarithms are given by
\beq \label{lemf}%\refeq{lemf}
\lemf:=\frac{\alpha}{4\pi}\left[\frac{1}{2}\log{\frac{\MW^2}{m_f^2}}+\log{\frac{\MW^2}{\la^2}}\right].
\eeq

\subsection*{Transverse physical gauge bosons $A,Z,W^\pm$}
\newcommand{\antikro}{E}
\newcommand{\kroAA}{\de_{V_a A}\de_{V_bA}}
\newcommand{\kroZZ}{\de_{\NB Z}\de_{\NB'Z}}
The collinear corrections for external physical gauge bosons
are related to the one-loop coefficients of the electroweak beta
functions. In the mass-eigenstate basis $V_a=A,Z,W^\pm$ these
coefficients generalize to a matrix $b^\ew_{ab}$ in the adjoint
representation (\cf appendix of \citere{DennPozz1}).  The charged
component reads
\beq
\bew_{W^\si W^{\si'}}=\de_{\si\si'}\frac{19}{6\sw^2},
\eeq
and determines the running of the $\SUtwo$ gauge coupling. In the neutral sector we have
\beqar \label{betarelations}%\refeq{betarelations}
\bew_{AA}&=&-\frac{11}{3},\qquad 
\bew_{AZ}=\bew_{ZA}=-\frac{19+22\sw^2}{6\sw\cw},\qquad
\bew_{ZZ}=\frac{19-38\sw^2-22\sw^4}{6\sw^2\cw^2}.
\eeqar
The $AA$ component determines the running of the electric
charge, and the $AZ$ component is associated with the running of the
weak mixing angle [\cf \refeq{counterterms}]. 
The SL corrections 
for transverse gauge bosons are given by
\beq \label{deccWT}%\refeq{deccWT}
\delta^\cc_{V_aV_b}(\GB_{\rT})=\frac{1}{2}\left[\bew_{V_aV_b}+\antikro_{V_aV_b}\bew_{AZ}\right]\ls +\de_{V_aV_b} Q_{V_a}^2\lemW-\frac{1}{2}\de_{V_a A}\de_{V_bA} \Delta \alpha (\MW^2).
\eeq
The first term corresponds to the result for a symmetric massless
gauge theory like QCD (see for instance \citere{Kunszt}). The second
term is proportional to the antisymmetric matrix $\antikro_{V_aV_b}$,
with non-vanishing components $E_{AZ}=-E_{ZA}=1$.  This term results
from the on-shell renormalization condition \cite{FortPhys} and
ensures that the correction factor for external photons does not
involve mixing with $\PZ$ bosons,
\beq
\de^\cc_{ZA}(\GB_{\rT})=0.
\eeq
The third term in \refeq{deccWT}
represents an electromagnetic contribution for charged external gauge bosons. Finally, the $AA$ component receives a pure electromagnetic contribution associated with the light-fermion loops,
\beq
\Delta \alpha (\MW^2)=
\frac{\al}{3\pi}\sum_{f,i,\si\neq t} \NCf Q^2_{f_\si}\log\frac{\MW^2}{m^2_{f_{\si,i}}}
\eeq
where the sum runs over the generations $i=1,2,3$ of leptons and
quarks $f=l,q$ with isospin $\si$ and colour factor $\NCf$, omitting
the top-quark contribution.

\subsection*{Longitudinally polarized gauge bosons $\PZ$, $\PW^\pm$}\label{loggaugebos}%\ref{loggaugebos}
The mass singular corrections for external longitudinal gauge bosons
$\PZ$, $\PW^\pm$, are obtained from the corrections for the
corresponding would-be Goldstone bosons $\chi$, $\phi^\pm$ using the
equivalence theorem. For renormalized amputated Green functions we
have the relations
\beqar \label{eq:renet} %\refeq{eq:renet}
p^\mu\langle W^\pm_{\mu}(p)\ldots\rangle&=&\pm\MW(1+\de C_{\phi})\langle \phi^\pm_0(p) \ldots\rangle,\nl
p^\mu\langle Z_{\mu}(p)\ldots\rangle&=&\ri\MZ(1+\de C_{\chi})\langle\chi_0(p) \ldots\rangle.
\eeqar
Besides the lowest-order contribution, \refeq{eq:renet} contains
non-trivial higher-order corrections owing to the mixing between gauge
bosons and would-be Goldstone bosons \cite{etcorr}. These corrections
correspond to the FRC's for would-be Goldstone bosons in \refeq{subllogfact2}, 
and combining them with the collinear factors for would-be Goldstone bosons
one obtains 
\beqar \label{longeq:coll} %\refeq{longeq:coll}
\de^\cc_{\phi^\pm\phi^\pm}(\Phi)&=& 
\de C_\phi+\de^{\coll}_{\phi^\pm\phi^\pm}(\Phi)
=  \left[2\cew_\Phi-\frac{3}{4\sw^2}\frac{\Mt^2}{\MW^2}\right]\ls   +Q_\PW^2\lemW ,\nl
\de^\cc_{\chi\chi}(\Phi)&=& 
\de C_\chi+\de^{\coll}_{\chi\chi}(\Phi)=
\left[2\cew_\Phi-\frac{3}{4\sw^2}\frac{\Mt^2}{\MW^2}\right]\ls.
\eeqar
The result is written in terms of the eigenvalue of $\cew$ for the
scalar doublet $\Phi$ and contains large Yukawa contributions.

\subsection*{Higgs bosons}
The SL corrections \refeq{subllogfact2} for Higgs bosons read
\beq
\de^{\cc}_{HH}(\Phi)= \left[2\cew_\Phi-\frac{3}{4\sw^2}\frac{\Mt^2}{\MW^2}\right]\ls.
\eeq
Note that up to pure electromagnetic contributions, longitudinal gauge
bosons and Higgs bosons receive the same collinear SL corrections.

\section{Logarithms connected to parameter renormalization}
\label{se:ren}%\ref{chargeren}
\renewcommand{\eff}{\mathrm{eff}}
\renewcommand{\gt}{g_{\Pt}}
\newcommand{\gH}{\la}
\newcommand{\rt}{h_{\Pt}}
\newcommand{\rH}{h_{\PH}}
The logarithms related to UV divergences
originate from the renormalization of the dimensionless parameters
\beq
e,\qquad \cw=\frac{\MW}{\MZ},\qquad \rt =\frac{\Mt}{\MW}, \qquad  \rH =\frac{\MH^2}{\MW^2},
\eeq
\ie the electric charge, the weak mixing angle, and the dimensionless
mass ratios related to the top-quark Yukawa coupling and to the scalar
self-coupling, respectively.
These SL corrections  %are those that
determine the running of the couplings, and in one-loop approximation they are obtained from the Born matrix element $\M_0=\M_0(e,\cw,\rt,\rH)$
in the high-energy limit by
\beq \label{PRtransf}%\refeq{PRtransf}
\de^\pre \M = \left. \frac{\de\M_0}{\de e}\de e 
+ \frac{\de\M_0}{\de\cw}\de\cw 
+ \frac{\de\M_0}{\de\rt}\de\rt 
+ \frac{\de\M_0}{\de\rH}\de\rH^{\eff} 
\, \right|_{\mu^2=s}
.
\eeq
The contribution from the tadpole renormalization to the
renormalization of the scalar self-coupling (\cf\citere{bfm}) is
included in the effective counterterm $\de \rH^\eff$.  In the on-shell
scheme (and in LA) the counterterms read
\beqar \label{counterterms}% \refeq{counterterms}
\frac{\delta \cw^2}{\cw^2}&=&\frac{\sw}{\cw}\bew_{AZ}\lu,\qquad
\frac{\de e^2}{e^2}=-\bew_{AA}\lu+\Delta \alpha (\MW^2),\nl
\frac{\de\rt}{\rt} &=& \left\{\frac{1}{2}\bew_{WW} 
-\frac{3}{2}\left(\cew_{\Pt^\rR}+\cew_{\Pt^\rL}\right)
+\frac{9}{8\sw^2}\frac{\Mt^2}{\MW^2}\right\}\lu,\nl
\frac{\de\rH^{\eff}}{\rH}&=& \left\{\bew_{WW} +
\frac{3}{2\sw^2}\left[\frac{\MW^2}{\MH^2}\left(2+\frac{1}{\cw^4}\right)
-\left(2+\frac{1}{\cw^2}\right) + \frac{\MH^2}{\MW^2}\right]\right.\nl
&&\left.{}
+\frac{3}{\sw^2}\frac{\Mt^2}{\MW^2}\left(1-2\frac{\Mt^2}{\MH^2}\right)\right\}\lu.
\eeqar
In the case of processes with longitudinal gauge bosons, the
renormalization \refeq{PRtransf} must be performed in the matrix
elements resulting from the equivalence theorem.

\section{Application to $\PW$-boson-pair production}
\label{se:applicat}%\ref{se:applicat}
In this section, the above results for Sudakov DL, collinear or soft
SL, and PR corrections are applied to $\PW$-pair production. Similar
results for neutral gauge-boson-pair production and neutral current
processes $\Pep\Pem\rightarrow f\bar{f}$ can be found in
\citere{DennPozz1}.
 
\newcommand{\NC}{R}
\newcommand{\NCew}{\NC_{\Pe^-_\kappa\PW^-_\la}}
\newcommand{\NCep}{\NC_{\Pe^-_\kappa\phi^-}}
\newcommand{\deNClq}{\Delta_{l^\kappa_\si q^\la_\rho}}
\newcommand{\deNCew}{\Delta_{\Pe^-_\kappa\PW^-_\rT}}
\newcommand{\deNCep}{\Delta_{\Pe^-_\kappa\phi^-}}
\newcommand{\NClq}{\NC_{l^\kappa_\si q^\la_\rho}}
\newcommand{\NCll}{\NC_{l^\kappa_\si l^\kappa_\si}}
\newcommand{\NCqq}{\NC_{q^\la_\rho q^\la_\rho}}
\newcommand{\NClmq}{\NC_{l^\kappa_{-\si} q^\la_\rho}}
\newcommand{\NClqm}{\NC_{l^\kappa_\si q^\la_{-\rho}}}

We consider the polarized scattering process%
\footnote{The momenta and
  fields of the initial states are incoming, and those of the
  final states are outgoing.}  
$\Pe^+_\kappa\Pe^-_\kappa \rightarrow \PW^+_{\la_+} \PW^-_{\la_-}$, where
$\kappa=\rR,\rL$ is the
electron chirality, and $\la_\pm=0,\pm$ represent the gauge-boson
helicities. In the high-energy limit only the following helicity
combinations are non-suppressed 
\cite{eeWWhe,Denn1}: the purely longitudinal final state
$(\la_+,\la_-)=(0,0)$, which we denote by $\PW^+_\rL\PW^-_\rL$, %$(\la_+,\la_-)=(\rL,\rL)$,
and the purely transverse and opposite final states  $(\la_+,\la_-)=(\pm,\mp)$,
which we denote by $\PW^+_\rT\PW^-_\rT$%$(\la_+,\la_-)=(\rT,\rT)$
. %All these final states,can be written as $(\la_+,\la_-)=(\la,-\la)$.
The Mandelstam
variables are $s=(p_\Pep+p_\Pem)^2$, $t=(p_\Pep-p_\PWp)^2\sim
-s(1-\cos{\theta})/2$, and $u=(p_\Pep-p_\PWm)^2\sim
-s(1+\cos{\theta})/2$, where $\theta$ is the angle between $\Pep$ and
$\PWp$.  The Born amplitude gets contributions of the $s$- and
$t$-channel diagrams in \reffi{WWborn} and reads
\beq \label{borneeww} %\refeq{borneeww}
\M_{0}^{ \Pe^+_\kappa\Pe^-_\kappa \rightarrow \PW^+_\rL\PW^-_\rL}=
e^2
\NCep
 \frac{A_s}{s},\qquad
\M_{0}^{\Pe^+_\kappa\Pe^-_\kappa \rightarrow
  \PW^+_\rT\PW^-_\rT}=\de_{\kappa\rL}\frac{e^2}{2\sw^2}\frac{A_t}{t} 
\eeq
up to terms of order $\MW^2/s$,  where 
$\NC_{\varphi_i\varphi_k}:=I^{A}_{\varphi_i}I^{A}_{\varphi_k}+I^{Z}_{\varphi_i}I^{Z}_{\varphi_k}$  is given by 
\beq
\NC_{\Pe^-_\rR\phi^-}=\frac{1}{2\cw^2},\quad
\NC_{\Pe^-_\rL\phi^-}=\frac{1}{4\sw^2\cw^2},\quad
\NC_{\Pe^-_\rL \PW^-_\rT}=\frac{1}{2\sw^2} .
\eeq
The amplitude for transverse gauge-boson production is 
is non-suppressed only for left-handed electrons in the initial state.
\begin{figure} [b!]
\begin{center}
\begin{picture}(300,80)
\put(0,0){
\begin{picture}(120,80)
\ArrowLine(20,30)(0,60)
\ArrowLine(0,0)(20,30)
\Vertex(20,30){2}
\Photon(20,30)(80,30){2}{6}
\Vertex(80,30){2}
\DashLine(100,0)(80,30){2}
\DashLine(80,30)(100,60){2}
\Text(-5,0)[r]{$\Pem$}
\Text(-5,60)[r]{$\Pep$}
\Text(105,0)[l]{$\phi^-$}
\Text(105,60)[l]{$\phi^+$}
\Text(50,35)[b]{$A,Z$}
\end{picture}}
\put(180,0){
\begin{picture}(120,80)
\ArrowLine(40,45)(10,60)
\ArrowLine(40,15)(40,45)
\ArrowLine(10,0)(40,15)
\Vertex(40,15){2}
\Vertex(40,45){2}
\Photon(70,0)(40,15){2}{3}
\Photon(40,45)(70,60){-2}{3}
\Text(5,0)[r]{$\Pem$}
\Text(5,60)[r]{$\Pep$}
\Text(47,30)[l]{$\nu_\Pe$}
\Text(75,0)[l]{$\PW^-_\rT$}
\Text(75,60)[l]{$\PW^+_\rT$}
\end{picture}}
\end{picture}
\end{center}
\caption{Dominant  lowest-order diagrams for
  $\Pe^+\Pe^-\to\phi^+\phi^-$ and $\Pe^+\Pe^-\to \PW^+_\rT \PW^-_\rT$}
\label{WWborn}
\end{figure}
In the following we give the one-loop corrections as relative corrections to the Born matrix elements \refeq{borneeww}.
The LSC contributions \refeq{SCsum} read
\beq\label{SCWW}
\de^{\SC}_{\Pe^+_\kappa\Pe^-_\kappa \rightarrow \PW^+_\la\PW^-_{-\la}}=-\sum_{\varphi=\Pe_\kappa,\PW_\la} \left[\cew_\varphi\Ls
+\Lemphi \right].
\eeq
Here and in the following formulas, the quantum numbers of the
would-be Goldstone bosons $\phi^\pm$ have to be used for longitudinally
polarized gauge bosons $\PW^\pm_\rL$. The eigenvalues of the effective
electroweak Casimir operator are
\beq
\cew_{\Pe_\rR}=\frac{1}{\cw^2},\quad
\cew_{\Pe_\rL}=\cew_{\Phi}=\frac{1+2\cw^2}{4\sw^2\cw^2},\quad
\cew_{\PW_\rT}=\frac{2}{\sw^2}.
\eeq
The $\SS$ corrections are obtained by applying \refeq{SScorr} to the
crossing symmetric process $\Pe^+_\kappa\Pe^-_\kappa
\PW^-_\la\PW^+_{-\la}\rightarrow 0$.  The contribution of the neutral
gauge bosons $V_a=A,Z$ gives
\beqar 
\sum_{V_a=A,Z}\de^{V_a,\SS}_{\Pe^+_\kappa\Pe^-_\kappa \rightarrow \PW^+_\la\PW^-_{-\la}}
=-\left[
4\NCew
\ls +
\frac{\alpha}{\pi}\log{\frac{\MW^2}{\la^2}}
\right] \ltu,
\eeqar
The contribution of soft $\PW^\pm$  bosons to \refeq{SScorr}
yields 
\beqar \label{eewwpmssc}%\refeq{eewwpmssc}
\sum_{\GB_a=W^\pm}\de^{\GB_a,\SS} \M^{\Pe^+_\kappa\Pe^-_\kappa\phi^-\phi^+}
&=&\frac{2\ls\de_{\kappa\rL}}{\sqrt{2}\sw}%\sum_{S=H,\chi}
\left[
\frac{1}{2\sw}\left(\M_0^{\bar{\nu}_\kappa\Pe^-_\kappa H \phi^+}
+\M_0^{\Pe^+_\kappa\nu_\kappa\phi^-H}\right)
\right.\nl&&\left.
{}+\frac{\ri}{2\sw} \left(\M_0^{\bar{\nu}_\kappa\Pe^-_\kappa \chi \phi^+}
-\M_0^{\Pe^+_\kappa\nu_\kappa\phi^-\chi}\right)
\right]\log{\frac{|t|}{s}}
,\nl
\sum_{\GB_a=W^\pm}\de^{\GB_a,\SS} \M^{\Pe^+_\rL\Pe^-_\rL\PW_\rT^-\PW_\rT^+}
&=&\frac{2\ls}{\sqrt{2}\sw}\left[
\left( \M_0^{\bar{\nu}_\rL\Pe^-_\rL A_\rT \PW_\rT^+}
+\M_0^{\Pe^+_\rL\nu_\rL\PW_\rT^-A_\rT}\right)
\right.\nl&&\left.
{}-\frac{\cw}{\sw}\left( \M_0^{\bar{\nu}_\rL\Pe^-_\rL Z_\rT \PW_\rT^+}
+\M_0^{\Pe^+_\rL\nu_\rL\PW_\rT^-Z_\rT}\right)
\right]
\log{\frac{|t|}{s}},
\eeqar
and after  explicit evaluation of the $\SUtwo$-transformed Born matrix elements on the left-hand side of \refeq{eewwpmssc}, we find the  relative corrections
\beqar
\sum_{\GB_a=W^\pm}\de^{\GB_a,\SS}_{\Pe^+_\kappa\Pe^-_\kappa\rightarrow \PW_\rL^+\PW_\rL^-}
&=&-\ls\frac{\de_{\kappa\rL}}{\sw^4 \NC_{\Pe^-_\rL\phi^-}}\log{\frac{|t|}{s}},\nl
\sum_{\GB_a=W^\pm}\de^{\GB_a,\SS}_{\Pe^+_\rL\Pe^-_\rL\rightarrow \PW_\rT^+\PW_\rT^-}
&=&-\frac{2}{\sw^2}\left(1-\frac{t}{u}\right)\ls\log{\frac{|t|}{s}}.
\eeqar 
The collinear and soft SL corrections can be read off from \refeq{deccfer},
\refeq{deccWT}, and \refeq{longeq:coll},
\beqar
\de^\cc_{\Pe^+_\kappa\Pe^-_\kappa \rightarrow \PW^+_\rL\PW^-_\rL} &=& \left[3 \cew_{\Pe_\kappa}+ 4\cew_\Phi\right]\lsl-\frac{3}{2\sw^2}\frac{m^2_t}{\MW^2}\lYuk +\sum_{\varphi=\Pe,\PW}2\lemphi,\nl
\de^\cc_{\Pe^+_\rL\Pe^-_\rL \rightarrow \PW^+_\rT\PW^-_\rT} &=& \left[3 \cew_{\Pe_\rL}+\bew_{WW}\right]\lsl+\sum_{\varphi=\Pe,\PW}2\lemphi.
\eeqar
Here the Yukawa and non-Yukawa $\ls$ terms have been denoted by $\lYuk$ and $\lsl$, respectively.
Note that the (large) Yukawa contributions occur only for longitudinal
gauge bosons.

The PR logarithms are obtained from the renormalization of
\refeq{borneeww}. The corresponding $\ls$ terms are denoted by 
 $\lpr$, and according to \refeq{counterterms} given by
\beqar\label{RGWW}
\de^\pre_{\Pe^+_\rR\Pe^-_\rR \rightarrow \PW^+_\rL\PW^-_\rL} &=&
-\left[\frac{\sw}{\cw}\bew_{AZ}
+\bew_{AA}\right]\lpr+
\Delta \alpha (\MW^2)
,\nl
\de^\pre_{\Pe^+_\rL\Pe^-_\rL \rightarrow \PW^+_\rL\PW^-_\rL} &=&-\left[(1-\frac{\cw^2}{\sw^2})\frac{\sw}{\cw}\bew_{AZ}
+\bew_{AA}\right]\lpr+
\Delta \alpha (\MW^2)
,\nl
\de^\pre_{\Pe^+_\rL\Pe^-_\rL \rightarrow \PW^+_\rT\PW^-_\rT}
&=&-\bew_{WW}\lpr+
\Delta \alpha (\MW^2).
\eeqar

\begin{figure}[b!]
\begin{center}%\centerline{
\setlength{\unitlength}{1cm}
\begin{picture}(10,8.3)
\put(0,0){\includegraphics{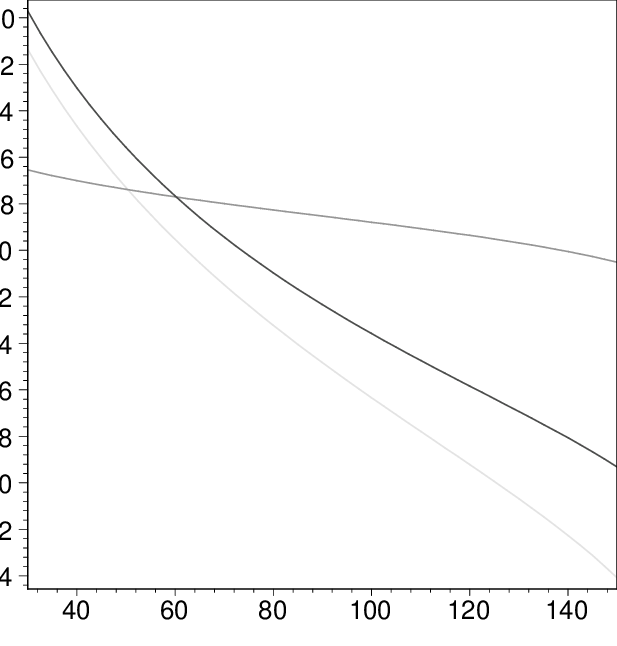}}
\put(2.5,0){\makebox(6,0.5)[b]{$\theta\,[^\circ]$}}
\put(-2.5,4){\makebox(1.5,1)[r]{$\delta^\sew\,[\%]$}}
\put(10,4.5){\makebox(1.5,1)[r]{$\rR\rL$}}
\put(10,2.0){\makebox(1.5,1)[r]{$\rL\rL$}}
\put(10,0.5){\makebox(1.5,1)[r]{$\rL\rT$}}
\end{picture}\end{center}%}
\caption[WWang]{Dependence of the electroweak correction factor
  $\de^{\sew}_{\Pe_\kappa^+\Pe_\kappa^-\rightarrow \PW_\la^+\PW_{-\la}^-}$ on
  the scattering angle $\theta$ at $\sqrt{s}=1\TeV$ for polarizations
  $\rR\rL$, $\rL\rL$, and $\rL\rT$} 
\label{plotWWan}
\end{figure}

\begin{figure}
\begin{center}%  \centerline{ 
\setlength{\unitlength}{1cm}
\begin{picture}(10,8.3)
\put(0,0){\includegraphics{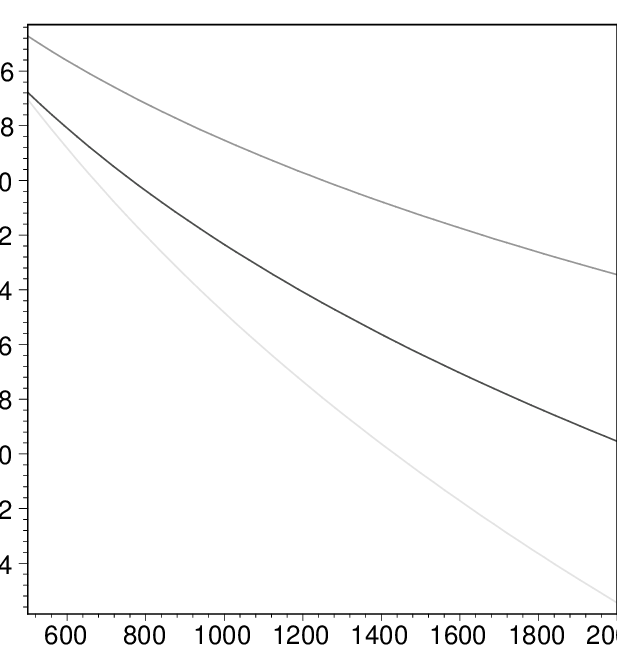}}
\put(2.5,0){\makebox(6,0.5)[b]{$\sqrt{s}\,[\GeV]$}}
\put(-2.5,4){\makebox(1.5,1)[r]{$\delta^\sew\,[\%]$}}
\put(10,4.5){\makebox(1.5,1)[r]{$\rR\rL$}}
\put(10,2.6){\makebox(1.5,1)[r]{$\rL\rL$}}
\put(10,0.8){\makebox(1.5,1)[r]{$\rL\rT$}}
\end{picture}\end{center}%}
\caption[WWang]{Dependence of the electroweak correction factor
  $\de^{\sew}_{\Pe_\kappa^+\Pe_\kappa^-\rightarrow \PW_\la^+\PW_{-\la}^-}$ on
  the energy  $\sqrt{s}$ at $\theta=90^\circ$ for polarizations
  $\rR\rL$, $\rL\rL$, and $\rL\rT$} 
\label{plotWWen}
\end{figure}

In order to give an impression of the size of the correction, we give a numerical evaluation of the symmetric electroweak part (ew) \refeq{dslogs} of the above results. 
Using the physical parameters
\beq
\MW=80.35 \GeV,\qquad \MZ=91.1867 \GeV,\qquad  \Mt=175 \GeV,\qquad
\alpha=\frac{1}{137.036},
\eeq
we obtain
\newcommand{\eeWWLL}{\de^\sew_{\Pe^+_\rL\Pe^-_\rL\rightarrow \PW^+_\rL\PW^-_\rL}}
\newcommand{\eeWWRL}{\de^{\sew}_{\Pe^+_\rR\Pe^-_\rR\rightarrow \PW^+_\rL\PW^-_\rL}}
\newcommand{\eeWWLT}{\de^{\sew}_{\Pe^+_\rL\Pe^-_\rL\rightarrow \PW^+_\rT\PW^-_\rT}}
\beqar
\eeWWLT &=&   -12.6\,\Ls-8.95\left[\ltu+\left(1-\frac{t}{u}\right)\lts\right]\ls
+25.2\,\lsl-14.2\,\lpr,\nl
\eeWWLL &=&   -7.35\,\Ls-\left(5.76\ltu+13.9\lts\right)\ls
+25.7\,\lsl-31.8\,\lYuk
\nl &&{}
-9.03\,\lpr   ,\nl
\eeWWRL &=&   -4.96\,\Ls-2.58\left(\ltu\right)\ls
{}+18.6\,\lsl-31.8\,\lYuk+8.80\,\lpr.
\eeqar 
These correction factors are shown in  \reffis{plotWWan} and
\ref{plotWWen} as a function of the scattering angle and the energy, respectively.
If the electrons are left-handed, large negative DL and PR
corrections originate from the $\SUtwo$ interaction. Instead, for
right-handed electrons the DL corrections are smaller, and the PR
contribution is positive. 
For transverse \PW~bosons, there are no Yukawa contributions and the
other contributions are in general larger than for longitudinal
\PW~bosons.  Nevertheless, for energies around $1\TeV$, the
corrections are similar.
Finally, note that the angular-dependent contributions are
very important for the LL and LT corrections: at $\sqrt{s}\approx 1\TeV$
they vary from $+15\%$ to $-5\%$ for scattering angles
$30^\circ<\theta<150^\circ$, whereas the angular-dependent part of the
RL corrections remains between $\pm 2\%$.


\section{Conclusion}

We have considered general electroweak processes at high energies.  We
have given recipes and explicit formulas for the extraction of the
one-loop leading electroweak logarithms. Like the well-known
soft--collinear double logarithms, also the collinear single
logarithms can be expressed as simple correction factors that are
associated with the external particles of the considered process. Up
to electromagnetic terms, the collinear SL corrections for external
longitudinal gauge bosons and for Higgs bosons are equal.  The
subleading single logarithms arising from the soft--collinear limit
are angular-dependent and can be associated to pairs of external
particles. Their evaluation requires in general all matrix elements
that are linked to the lowest-order matrix element via global $\SU(2)$
rotations.  Finally, the logarithms originating from coupling-constant
renormalization are associated with the explicit dependence of the
lowest-order matrix element on the coupling parameters.  
Our results
are applicable to general amplitudes that are not mass-suppressed, as
long as all invariants are large compared to the masses.  As
illustration, we have applied our general results to W-boson-pair
production.

\Acknowledgments
A.D. would like to thank the organizers of the conference, in particular
Howard Haber, for providing a very pleasant atmosphere during the
whole conference and for financial support.
This work was supported in part by the Swiss Bundesamt f\"ur Bildung und
Wissenschaft and by the European Union under contract
HPRN-CT-2000-00149.

 \newcommand{\vj}[4]{{\sl #1~}{\bf #2~}\ifnum#3<100 (19#3) \else (#3) \fi #4}
 \newcommand{\ej}[3]{{\bf #1~}\ifnum#2<100 (19#2) \else (#2) \fi #3}
 \newcommand{\vjs}[2]{{\sl #1~}{\bf #2}}

 \newcommand{\am}[3]{\vj{Ann.~Math.}{#1}{#2}{#3}}
 \newcommand{\ap}[3]{\vj{Ann.~Phys.}{#1}{#2}{#3}}
 \newcommand{\app}[3]{\vj{Acta~Phys.~Pol.}{#1}{#2}{#3}}
 \newcommand{\cmp}[3]{\vj{Commun. Math. Phys.}{#1}{#2}{#3}}
 \newcommand{\cnpp}[3]{\vj{Comments Nucl. Part. Phys.}{#1}{#2}{#3}}
 \newcommand{\cpc}[3]{\vj{Comp. Phys. Commun.}{#1}{#2}{#3}}
 \newcommand{\epj}[3]{\vj{Eur. Phys. J.}{#1}{#2}{#3}}
 \newcommand{\fp}[3]{\vj{Fortschr. Phys.}{#1}{#2}{#3}}
 \newcommand{\hpa}[3]{\vj{Helv. Phys.~Acta}{#1}{#2}{#3}}
 \newcommand{\ijmp}[3]{\vj{Int. J. Mod. Phys.}{#1}{#2}{#3}}
 \newcommand{\jetp}[3]{\vj{JETP}{#1}{#2}{#3}}
 \newcommand{\jetpl}[3]{\vj{JETP Lett.}{#1}{#2}{#3}}
 \newcommand{\jmp}[3]{\vj{J.~Math. Phys.}{#1}{#2}{#3}}
 \newcommand{\jp}[3]{\vj{J.~Phys.}{#1}{#2}{#3}}
 \newcommand{\lnc}[3]{\vj{Lett. Nuovo Cimento}{#1}{#2}{#3}}
 \newcommand{\mpl}[3]{\vj{Mod. Phys. Lett.}{#1}{#2}{#3}}
 \newcommand{\nc}[3]{\vj{Nuovo Cimento}{#1}{#2}{#3}}
 \newcommand{\nim}[3]{\vj{Nucl. Instr. Meth.}{#1}{#2}{#3}}
 \newcommand{\np}[3]{\vj{Nucl. Phys.}{#1}{#2}{#3}}
 \newcommand{\npbps}[3]{\vj{Nucl. Phys. B (Proc. Suppl.)}{#1}{#2}{#3}}
 \newcommand{\pl}[3]{\vj{Phys. Lett.}{#1}{#2}{#3}}
 \newcommand{\prp}[3]{\vj{Phys. Rep.}{#1}{#2}{#3}}
 \newcommand{\pr}[3]{\vj{Phys.~Rev.}{#1}{#2}{#3}}
 \newcommand{\prl}[3]{\vj{Phys. Rev. Lett.}{#1}{#2}{#3}}                       
 \newcommand{\ptp}[3]{\vj{Prog. Theor. Phys.}{#1}{#2}{#3}}                     
 \newcommand{\rpp}[3]{\vj{Rep. Prog. Phys.}{#1}{#2}{#3}}                       
 \newcommand{\rmp}[3]{\vj{Rev. Mod. Phys.}{#1}{#2}{#3}}                        
 \newcommand{\rnc}[3]{\vj{Revista del Nuovo Cim.}{#1}{#2}{#3}}                 
 \newcommand{\sjnp}[3]{\vj{Sov. J. Nucl. Phys.}{#1}{#2}{#3}}                   
 \newcommand{\sptp}[3]{\vj{Suppl. Prog. Theor. Phys.}{#1}{#2}{#3}}             
 \newcommand{\zp}[3]{\vj{Z. Phys.}{#1}{#2}{#3}}                                
 \renewcommand{\and}{and~}

\newpage


\begin{thebibliography}{99} 

\bibitem{cernreport} % \cite{cernreport}
S. Haywood, P.R. Hobson, W. Hollik, Z. Kunszt et al.,
hep-ph/0003275, in {\sl Standard Model Physics (and more) at the LHC},
eds.~G. Altarelli and M.L. Mangano, (CERN-2000-004, Gen\`eve, 2000) p.~117.
%%CITATION = HEP-PH 0003275;%%

\bibitem{LC1} % \cite{LC1}
E. Accomando et al., \prp{299}{98}{1} [hep-ph/9705442].
%%CITATION = HEP-PH 9705442;%%

\bibitem{Kuroda}%\cite{Kuroda}
M. Kuroda, G. Moultaka \and D. Schildknecht, \np{B350}{91}{25};\\
%%CITATION = NUPHA,B350,25;%%
G. Degrassi \and A. Sirlin, \pr{D46}{92}{3104};\\
%%CITATION = PHRVA,D46,3104;%%
A.~Denner, S. Dittmaier \and R. Schuster,
\np{B452}{95}{80} [hep-ph/9503442];\\
%%CITATION = HEP-PH 9503442;%%
A.~Denner, S. Dittmaier \and T. Hahn,
\pr{D56}{97}{117} [hep-ph/9612390].
%%CITATION = HEP-PH 9612390;%%%

\bibitem{eeWWhe}% \cite{eeWWhe}
W.~Beenakker et al.,
\np{B410}{93}{245} and \pl{B317}{93}{622}.
%%CITATION = NUPHA,B410,245;%%
%%CITATION = PHLTA,B317,622;%%

\bibitem{SUD}% \cite{SUD}
V.V. Sudakov, \jetp{3}{56}{65}.
%%CITATION = SPHJA,3,65;%%

\bibitem{CC0} %  \cite{CC0}
P. Ciafaloni \and D. Comelli, \pl{B446}{99}{278} [hep-ph/9809321].
%%CITATION = HEP-PH 9809321;%%


\bibitem{Ku1} % \cite{Ku1}
J.H. K\"uhn \and A.A. Penin, TTP-99-28, hep-ph/9906545.
%%CITATION = HEP-PH 9906545;%%

\bibitem{CC1} %\cite{CC1}
P. Ciafaloni \and D. Comelli, \pl{B476}{2000}{49} [hep-ph/9910278].
%%CITATION = HEP-PH 9910278;%%


\bibitem{Ku2} % \cite{Ku2}
J.H. K\"uhn, A.A. Penin \and V.A. Smirnov,
Eur.\ Phys.\ J.\ {\bf C17} (2000) 97
[hep-ph/9912503].
%%CITATION = HEP-PH 9912503;%%

\bibitem{Fa00}%
V.S. Fadin, L.N. Lipatov, A.D. Martin \and M. Melles,
\pr{D61}{2000}{094002}
[hep-ph/9910338].
%%CITATION = HEP-PH 9910338;%%


\bibitem{Me2} % \cite{Me2}
M. Melles,
Phys.\ Lett.\ {\bf B495} (2000) 81
[hep-ph/0006077].
%%CITATION = HEP-PH 0006077;%%

\bibitem{Be00}%
W. Beenakker \and A. Werthenbach,
\pl{B489}{2000}{148}
[hep-ph/0005316].
%%CITATION = HEP-PH 0005316;%%

\bibitem{Ho00}%
M. Hori, H. Kawamura \and J. Kodaira,
Phys.\ Lett.\ {\bf B491} (2000) 275
[hep-ph/0007329].
%%CITATION = HEP-PH 0007329;%%

\bibitem{be1}% \cite{be1}
  M. Beccaria, P. Ciafaloni, D. Comelli, F. Renard \and C.
  Verzegnassi, \pr{D61}{2000}{073005} and {011301} [hep-ph/9906319 and hep-ph/9907
389];\\
%%CITATION = HEP-PH 9906319;%%
%%CITATION = HEP-PH 9907389;%%
M. Beccaria, F.M. Renard \and C. Verzegnassi, PM-00-23, hep-ph/0007224.
%%CITATION = HEP-PH 0007224;%%

\bibitem{Me1} % \cite{Me1}
M. Melles,
Phys.\ Rev.\ D {\bf 63} (2001) 034003
[hep-ph/0004056].
%%CITATION = HEP-PH 0004056;%%

\bibitem{DennPozz1} % \cite{DennPozz1}
A.~Denner \and S.~Pozzorini,
Eur.\ Phys.\ J.\ C {\bf 18} (2001) 461
[hep-ph/0010201].
%%CITATION = HEP-PH 0010201;%%

\bibitem{FortPhys}% \cite{FortPhys}
A. Denner, \fp{41}{93}{307}.
%%CITATION = FPYKA,41,307;%%

\bibitem{KLN} % \cite{KLN}
T. Kinoshita, \jmp{3}{62}{650}.
%%CITATION = JMAPA,3,650;%%

\bibitem{DennPozz2} % \cite{DennPozz2}
A.~Denner and S.~Pozzorini,
%``One-loop leading logarithms in electroweak radiative corrections.  II:
%Factorization of collinear singularities,''
Eur.\ Phys.\ J.\ C {\bf 21} (2001) 63
[hep-ph/0104127].
%%CITATION = HEP-PH 0104127;%%

\bibitem{et}%
J.M. Cornwall, D.N. Levin \and G. Tiktopoulos, \pr{D10}{74}{1145}; \\
%%CITATION = PHRVA,D10,1145;%%
G.J. Gounaris, R. K\"ogerler \and H. Neufeld, \pr{D34}{86}{3257}.
%%CITATION = PHRVA,D34,3257;%%

\bibitem{etcorr} %21 \cite{etcorr}
Y.P. Yao \and C.P. Yuan, \pr{D38}{88}{2237};\\
%%CITATION = PHRVA,D38,2237;%%
J. Bagger \and C. Schmidt, \pr{D41}{90}{264};\\
%%CITATION = PHRVA,D41,264;%%
H.J. He, Y.P. Kuang \and X. Li, \prl{69}{92}{2619} and \pr{D49}{94}{4842};\\
%%CITATION = PRLTA,69,2619;%%
%%CITATION = PHRVA,D49,4842;%%
D. Espriu \and J. Matias, \pr{D52}{95}{6530}
[hep-ph/9501279].
%%CITATION = HEP-PH 9501279;%%

\bibitem{Kunszt}%\cite{Kunszt}
Z. Kunszt, A. Signer \and Z. Trocsanyi, \np{B420}{94}{550}
[hep-ph/9401294].
%%CITATION = HEP-PH 9401294;%%

\bibitem{bfm}%
A. Denner, S. Dittmaier \and G. Weiglein, \np{B440}{95}{95}
[hep-ph/9410338].
%%CITATION = HEP-PH 9410338;%%

\bibitem{Denn1} % \cite{Denn1}
W. Beenakker \and A. Denner, \ijmp{A9}{94}{4837}.
%%CITATION = IMPAE,A9,4837;%%


\end{thebibliography}
\end{document}